\input harvmac
\input epsf
\def\ev#1{\langle#1\rangle}

\noblackbox
\def\IL{\relax{\rm I\kern-.18em L}}
\def\IH{\relax{\rm I\kern-.18em H}}
\def\IR{\relax{\rm I\kern-.18em R}}
\def\IC{\relax\hbox{$\inbar\kern-.3em{\rm C}$}}
\def\IZ{\relax\ifmmode\mathchoice
{\hbox{\cmss Z\kern-.4em Z}}{\hbox{\cmss Z\kern-.4em Z}}
{\lower.9pt\hbox{\cmsss Z\kern-.4em Z}} {\lower1.2pt\hbox{\cmsss
Z\kern-.4em Z}}\else{\cmss Z\kern-.4em Z}\fi}

\def\CJ {{\cal J}}

\def\CL {{\cal L}}
\def\CV {{\cal V}}
\def\CO {{\cal O}}


\font\manual=manfnt \def\dbend{\lower3.5pt\hbox{\manual\char127}}

\def\IZ{\relax\ifmmode\mathchoice
{\hbox{\cmss Z\kern-.4em Z}}{\hbox{\cmss Z\kern-.4em Z}}
{\lower.9pt\hbox{\cmsss Z\kern-.4em Z}} {\lower1.2pt\hbox{\cmsss
Z\kern-.4em Z}}\else{\cmss Z\kern-.4em Z}\fi}
\def\half {{1\over 2}}

\font\small =cmr10 scaled 850
\def\vev#1{\langle#1\rangle}
\def\frac#1#2{{{#1}\over{#2}}}
\def\half{\frac{1}{2}}
\def\centertable#1{ \hbox to \hsize {\hfill\vbox{
                    \offinterlineskip \tabskip=0pt \halign{#1} }\hfill} }

\def\wt{\widetilde}
\def\Box{\hbox{$\rlap{$\sqcap$}\sqcup$}}

\def\J{{\cal J}}
\vskip-.2in
\lref\aglr{
  N.~Arkani-Hamed, G.~F.~Giudice, M.~A.~Luty and R.~Rattazzi,
  ``Supersymmetry-breaking loops from analytic continuation into  superspace,''
  Phys.\ Rev.\  D {\bf 58}, 115005 (1998)
  [arXiv:hep-ph/9803290].
}
\lref\mss{
  P.~Meade, N.~Seiberg and D.~Shih,
  ``General Gauge Mediation,''
  arXiv:0801.3278 [hep-ph].
}
\lref\ADS{
  I.~Affleck, M.~Dine and N.~Seiberg,
  ``Dynamical Supersymmetry Breaking In Four-Dimensions And Its
  Phenomenological Implications,''
  Nucl.\ Phys.\  B {\bf 256}, 557 (1985).
}
\lref\GiudiceBP{
  G.~F.~Giudice and R.~Rattazzi,
  ``Theories with gauge-mediated supersymmetry breaking,''
  Phys.\ Rept.\  {\bf 322}, 419 (1999)
  [arXiv:hep-ph/9801271].
}
\lref\sv{
  M.~A.~Shifman and A.~I.~Vainshtein,
  ``Solution of the Anomaly Puzzle in SUSY Gauge Theories and the Wilson
  Operator Expansion,''
  Nucl.\ Phys.\  B {\bf 277}, 456 (1986)
  [Sov.\ Phys.\ JETP {\bf 64}, 428 (1986\ ZETFA,91,723-744.1986)].
}
\lref\DistlerBT{
  J.~Distler and D.~Robbins,
  ``General F-Term Gauge Mediation,''
  arXiv:0807.2006 [hep-ph].
}
\lref\OoguriEZ{
  H.~Ooguri, Y.~Ookouchi, C.~S.~Park and J.~Song,
  ``Current Correlators for General Gauge Mediation,''
  arXiv:0806.4733 [hep-th].
}

\lref\CarpenterWI{
  L.~M.~Carpenter, M.~Dine, G.~Festuccia and J.~D.~Mason,
  ``Implementing General Gauge Mediation,''
  arXiv:0805.2944 [hep-ph].
}
\lref\EndoGI{
  M.~Endo and K.~Yoshioka,
  ``Low-scale Gaugino Mass Unification,''
  arXiv:0804.4192 [hep-ph].
}
\lref\IbeSI{
  M.~Ibe, Y.~Nakayama and T.~T.~Yanagida,
  ``Conformal Gauge Mediation and Light Gravitino of Mass $m_{3/2} <O(10eV)$,"
arXiv:0804.0636 [hep-ph].
}
\lref\IbeAB{
  M.~Ibe and R.~Kitano,
  ``Minimal Direct Gauge Mediation,''
  Phys.\ Rev.\  D {\bf 77}, 075003 (2008)
  [arXiv:0711.0416 [hep-ph]].
}
\lref\MartinVX{
  S.~P.~Martin,
  ``Two-loop effective potential for a general renormalizable theory and
  softly broken supersymmetry,''
  Phys.\ Rev.\  D {\bf 65}, 116003 (2002)
  [arXiv:hep-ph/0111209].
}

\lref\WessCP{
  J.~Wess and J.~Bagger,
  ``Supersymmetry and supergravity,''
{\it  Princeton, USA: Univ. Pr. (1992) 259 p}
}
\lref\os{
  H.~Osborn,
  ``N = 1 superconformal symmetry in four-dimensional quantum field theory,''
  Annals Phys.\  {\bf 272}, 243 (1999)
  [arXiv:hep-th/9808041].
}
\lref\gr{
  G.~F.~Giudice and R.~Rattazzi,
  ``Extracting supersymmetry-breaking effects from wave-function
  renormalization,''
  Nucl.\ Phys.\  B {\bf 511}, 25 (1998)
  [arXiv:hep-ph/9706540].
}
\lref\dMM{
  A.~de Gouvea, T.~Moroi and H.~Murayama,
  ``Cosmology of supersymmetric models with low-energy gauge mediation,''
  Phys.\ Rev.\  D {\bf 56}, 1281 (1997)
  [arXiv:hep-ph/9701244].
}
\lref\egms{
  E.~Gorbatov and M.~Sudano,
  ``Sparticle Masses in Higgsed Gauge Mediation,''
  arXiv:0802.0555 [hep-ph].
}
\lref\DistlerBT{
  J.~Distler and D.~Robbins,
  ``General F-Term Gauge Mediation,''
  arXiv:0807.2006 [hep-ph].
}
\lref\DineVC{
  M.~Dine, A.~E.~Nelson and Y.~Shirman,
  ``Low-Energy Dynamical Supersymmetry Breaking Simplified,''
  Phys.\ Rev.\  D {\bf 51}, 1362 (1995)
  [arXiv:hep-ph/9408384].
}
\lref\DineAG{
  M.~Dine, A.~E.~Nelson, Y.~Nir and Y.~Shirman,
  ``New tools for low-energy dynamical supersymmetry breaking,''
  Phys.\ Rev.\  D {\bf 53}, 2658 (1996)
  [arXiv:hep-ph/9507378].
}
\lref\CohenQC{
  A.~G.~Cohen, T.~S.~Roy and M.~Schmaltz,
  ``Hidden sector renormalization of MSSM scalar masses,''
  JHEP {\bf 0702}, 027 (2007)
  [arXiv:hep-ph/0612100].
}
\lref\MurayamaGE{
  H.~Murayama, Y.~Nomura and D.~Poland,
  ``More Visible Effects of the Hidden Sector,''
  Phys.\ Rev.\  D {\bf 77}, 015005 (2008)
  [arXiv:0709.0775 [hep-ph]].
}
\lref\DineYW{
  M.~Dine and A.~E.~Nelson,
  ``Dynamical supersymmetry breaking at low-energies,''
  Phys.\ Rev.\  D {\bf 48}, 1277 (1993)
  [arXiv:hep-ph/9303230].
}
\lref\RoyNZ{
  T.~S.~Roy and M.~Schmaltz,
  ``A hidden solution to the $\mu/B_\mu$ problem in gauge mediation,''
  Phys.\ Rev.\  D {\bf 77}, 095008 (2008)
  [arXiv:0708.3593 [hep-ph]].
}
\lref\oops{
  H.~Ooguri, Y.~Ookouchi, C.~S.~Park and J.~Song,
  ``Current Correlators for General Gauge Mediation,''
  arXiv:0806.4733 [hep-th].
}
\lref\LutyFK{
  M.~A.~Luty,
  ``Naive dimensional analysis and supersymmetry,''
  Phys.\ Rev.\  D {\bf 57}, 1531 (1998)
  [arXiv:hep-ph/9706235].
}
\lref\DineXK{
  M.~Dine, Y.~Nir and Y.~Shirman,
  ``Variations on minimal gauge mediated supersymmetry breaking,''
  Phys.\ Rev.\  D {\bf 55}, 1501 (1997)
  [arXiv:hep-ph/9607397].
}
\lref\DineZA{
  M.~Dine, W.~Fischler and M.~Srednicki,
  ``Supersymmetric Technicolor,''
  Nucl.\ Phys.\  B {\bf 189}, 575 (1981).
}
\lref\DimAU{
  S.~Dimopoulos and S.~Raby,
  ``Supercolor,''
  Nucl.\ Phys.\  B {\bf 192}, 353 (1981).
}
\lref\DineGU{
  M.~Dine and W.~Fischler,
  ``A Phenomenological Model Of Particle Physics Based On Supersymmetry,''
  Phys.\ Lett.\  B {\bf 110}, 227 (1982).
}
\lref\NappiHM{
  C.~R.~Nappi and B.~A.~Ovrut,
  ``Supersymmetric Extension Of The SU(3) X SU(2) X U(1) Model,''
  Phys.\ Lett.\  B {\bf 113}, 175 (1982).
}
\lref\Alvarez{
  L.~Alvarez-Gaume, M.~Claudson and M.~B.~Wise,
  ``Low-Energy Supersymmetry,''
  Nucl.\ Phys.\  B {\bf 207}, 96 (1982).
}
\lref\DimGM{
  S.~Dimopoulos and S.~Raby,
  ``Geometric Hierarchy,''
  Nucl.\ Phys.\  B {\bf 219}, 479 (1983).
}
\def\gmed{\refs{\DineZA\DimAU\NappiHM\Alvarez\DimGM\ADS\DineYW\DineVC\DineAG\DineXK-\GiudiceBP}}
\Title{\vbox{\baselineskip12pt \hbox{UCSD-PTH-08-05}}}
{\vbox{\centerline{Comments on General Gauge Mediation}}}
\smallskip
\centerline{Kenneth Intriligator and Matthew Sudano}
\smallskip
\bigskip
\centerline{{\it Department of Physics, University of
California, San Diego, La Jolla, CA 92093 USA}}
\bigskip
\vskip 1cm

\noindent 
There has been interest in generalizing models of gauge mediation of supersymmetry breaking.  As shown by Meade, Seiberg, and Shih (MSS), the soft masses of general gauge mediation can be expressed in terms of the current two-point functions of the susy-breaking sector.  We here give a simple extension of their result which provides, for general gauge mediation, the full effective potential for squark pseudo-D-flat directions.  The effective potential reduces to the sfermion soft masses near the origin, and the full potential, away from the origin, can be useful for cosmological applications.  We also generalize the soft masses and effective potential to allow for general gauge mediation by Higgsed gauge groups. 
Finally, we discuss general gauge mediation in the limit of small F-terms, and how the results of MSS connect with the analytic continuation in superspace results, based on a spurion analysis.

\bigskip

\Date{July 2008}

\newsec{Introduction}
A standard framework for building potentially realistic supersymmetric models is based on theories of the form
\eqn\lsum{\CL = \CL _{1}+\CL _{2}+\CL _{int},}
where $\CL _{1}$ is the MSSM or some extension, $\CL _2$ is the hidden sector with broken supersymmetry, and the $\CL _{int}$ interactions couple them.  In gauge mediation \gmed,
the gauge interactions are the most important part of $\CL _{int}$.  This scenario has been extensively studied for simple, weakly coupled, hidden sectors $\CL _2$.  It is potentially interesting to extend such results to more complicated hidden sectors, including those which are not necessarily weakly coupled, see e.g.  \refs{\ADS,  \LutyFK\CohenQC\RoyNZ-\MurayamaGE}.  A general framework that can accommodate this scenario was considered in \mss, where it was shown that the soft masses of MSSM gauginos and sfermions, to leading order in the $\CL _{int}$ gauge interactions, can be expressed in terms of the $\CL _2$ current correlation functions.   
Related following works include \refs{\IbeAB \IbeSI \EndoGI\CarpenterWI\OoguriEZ-\DistlerBT}.

In this short note, we extend the results of \mss\ to compute the full effective potential for the sfermion fields.  Expanding the effective potential around the origin gives the sfermion masses.  The form of the effective potential far from the origin can be of interest for cosmological models, as in \dMM.  When the susy-breaking/messenger sector $\CL _2$ is weakly coupled and expanded for small susy-breaking F-terms,  our general effective potential reduces to that obtained in \dMM.   We also express the full effective potential, generalized to allow for the possibility that the messenger gauge group is Higgsed.   Expanding around the origin, this gives the gaugino and sfermion masses for general Higgsed gauge mediation.  When the susy-breaking sector $\CL _2$ is weakly coupled, these results reduce to those recently obtained in \egms\ for Higgsed gauge mediation.  
Finally, we discuss a relation between these current-correlator results and the discussion in \aglr\ of the 1PI effective action and RG running.  We discuss the results of \mss\ in terms of superspace and show how a spurion analysis reproduces the results obtained by analytic continuation in superspace \refs{\gr, \aglr} in the limit of small supersymmetry breaking.

The organization of this paper is as follows.  In section 2, we give a brief review of General Gauge Mediation \mss.  In section 3, the full effective potential is presented in this formalism, and the generalization to Higgsed gauge groups is discussed.  In section 4, superspace techniques \refs{\gr, \aglr} are used to extract results for small F-term breaking.  The main observations of section 4 were independently obtained in the recent work  \DistlerBT 

\newsec{Review of General Gauge Mediation \mss}
In supersymmetric gauge theories, the gauge supermultiplet $\CV$ couples to the current 
supermultiplet, $\CJ$, which is a real linear superfield satisfying 
$D^2\J=\bar D^2\J=0$.  In components, 
\eqn\superj{\J=J+i\theta j-i\bar\theta\bar j-\theta\sigma^\mu\bar\theta j_\mu+\half\theta\theta\bar\theta\bar\sigma^\mu\partial_\mu j-\half\bar\theta\bar\theta\theta\sigma^\mu\partial_\mu\bar j-\frac{1}{4}\theta\theta\bar\theta\bar\theta\Box J,}
with $\partial ^\mu j_\mu=0$ and the other components unconstrained.  The gauge interactions couple to $\CJ$ as 
\eqn\lintcont{\CL _{int}\supset 2g\int d^4 \theta \CJ\CV+\dots=g(JD-\lambda j-\bar \lambda \bar j -j^\mu V_\mu)+\dots,}
where the component expansion is in  Wess-Zumino gauge.  As shown in \mss,  the diagrams of Figure 1, which give the soft supersymmetry breaking masses of the visible sector,  can be expressed in terms of the hidden-sector current-current two-point functions.  Lorentz invariance and current conservation fix the form of the Euclidean momentum-space two-point functions of these fields as (dropping a $(2\pi )^4\delta ^{(4)}(0)$):
\eqna\compjj
$$\eqalignno{
&\vev{J(p)J(-p)}=\wt C_0(p^2/M^2)&\compjj a\cr
&\vev{j_\alpha(p)\bar j_{\dot\alpha}(-p)}=-\sigma_{\alpha\dot\alpha}^\mu p_\mu\wt C_{1/2}(p^2/M^2)&\compjj b\cr
&\vev{j_\mu(p)j_\nu(-p)}=-(p^2\eta_{\mu\nu}-p_\mu p_\nu)\wt C_1(p^2/M^2)&\compjj c\cr
&\vev{j_\alpha(p)j_\beta(-p)}=\epsilon_{\alpha\beta}M\wt B_{1/2}(p^2/M^2)&\compjj d
}$$
for some functions, $\wt C_0$, $\wt C_{1/2}$, $\wt C_1$, and $\wt B_{1/2}$.  If supersymmetry were unbroken, $\widetilde C_0=\widetilde C_{1/2}=\widetilde C_1$, and $\widetilde B_{1/2} =0$.  Here $M$ is a mass scale in the problem.   We are interested in these two-point functions in the hidden sector, where supersymmetry is broken.   The $\widetilde C_{j=0, 1/2, 1}$ also depend on a UV cutoff, which is needed to regulate the Fourier transform from position space to momentum space, as
\eqn\ctildejlam{\widetilde C_j(p^2/M^2)=2\pi ^2 c \log (\Lambda/M)+\widetilde C_j^{finite}(p^2/M^2)}
where only the finite terms $\wt C_j^{finite}$ depend on $j$ when supersymmetry is spontaneously broken.  Also, in this case, $\widetilde B_{1/2}$ is independent of $\Lambda$ \mss.    

To $\CO(g^2)$, the hidden sector then contributes to the effective action for the gauge supermultiplet fields as \mss\ 
\eqn\gaugecont{\eqalign{\delta \CL _{eff}&=\half g^2 \widetilde C_0(0)D^2-g^2 \widetilde C_{1/2}(0)i\lambda \sigma ^\mu \partial _\mu \bar \lambda -{1\over 4}g^2 \widetilde C_1(0)F_{\mu \nu}F^{\mu \nu}\cr &-{1\over 2} g^2 (M\widetilde B_{1/2}(0)\lambda \lambda + c.c. )+\dots.}}
\def\size{2}
\vskip-.1in
\centertable{
\vrule height.3ex depth0.25ex width 0pt \tabskip=1em  \hfil#\hfil\cr
\epsfxsize=\size truein\epsfbox{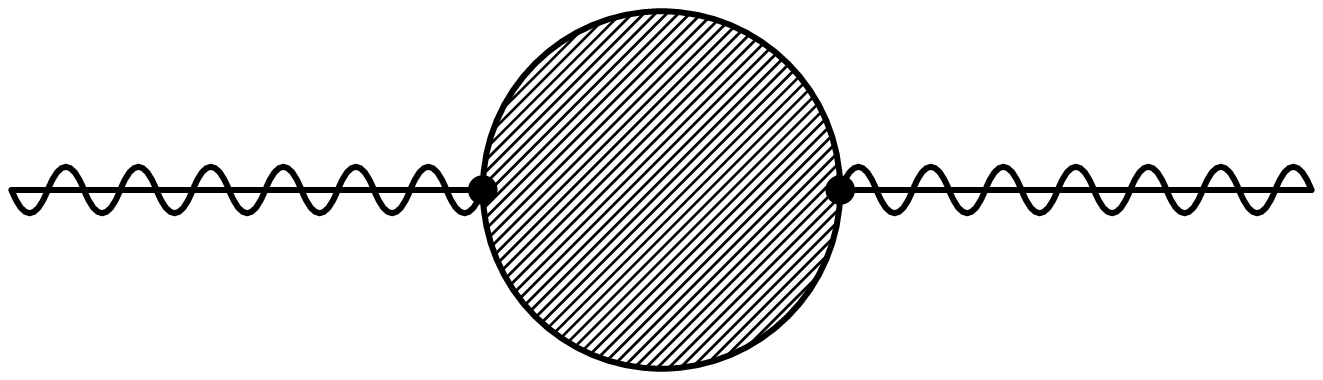}\cr
\cr
$D1$\cr
}
\vskip.2in
\centertable{
\vrule height.3ex depth0.25ex width 0pt \tabskip=1em  \hfil#\hfil\qquad&\qquad\hfil#\hfil\cr
\epsfxsize=\size truein\epsfbox{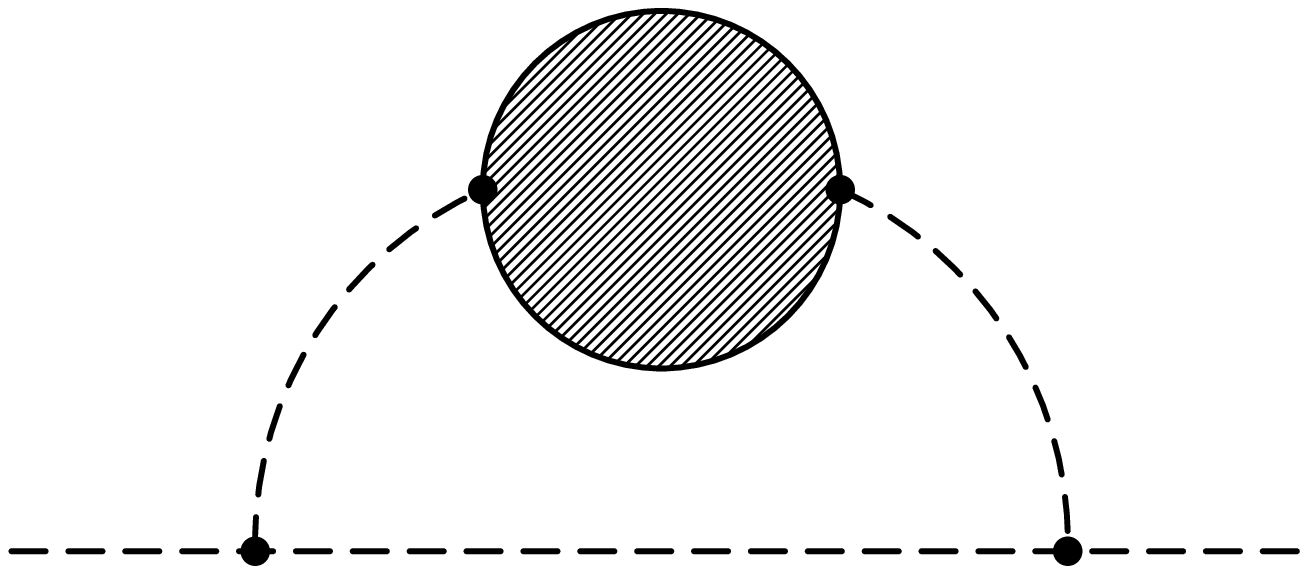}&\epsfxsize=\size truein\epsfbox{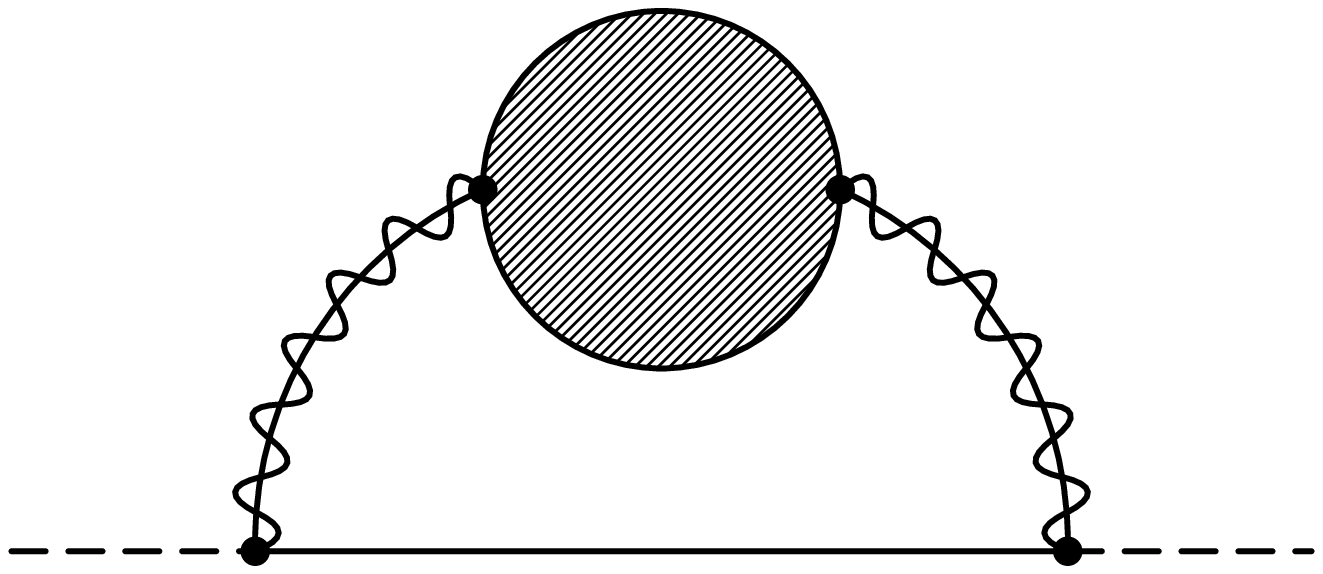}\cr
&\cr
$D2$&$D3$\cr
}
\vskip.2in
\centertable{
\vrule height.3ex depth0.25ex width 0pt \tabskip=1em  \hfil#\hfil\qquad&\qquad\hfil#\hfil\cr
\epsfxsize=\size truein\epsfbox{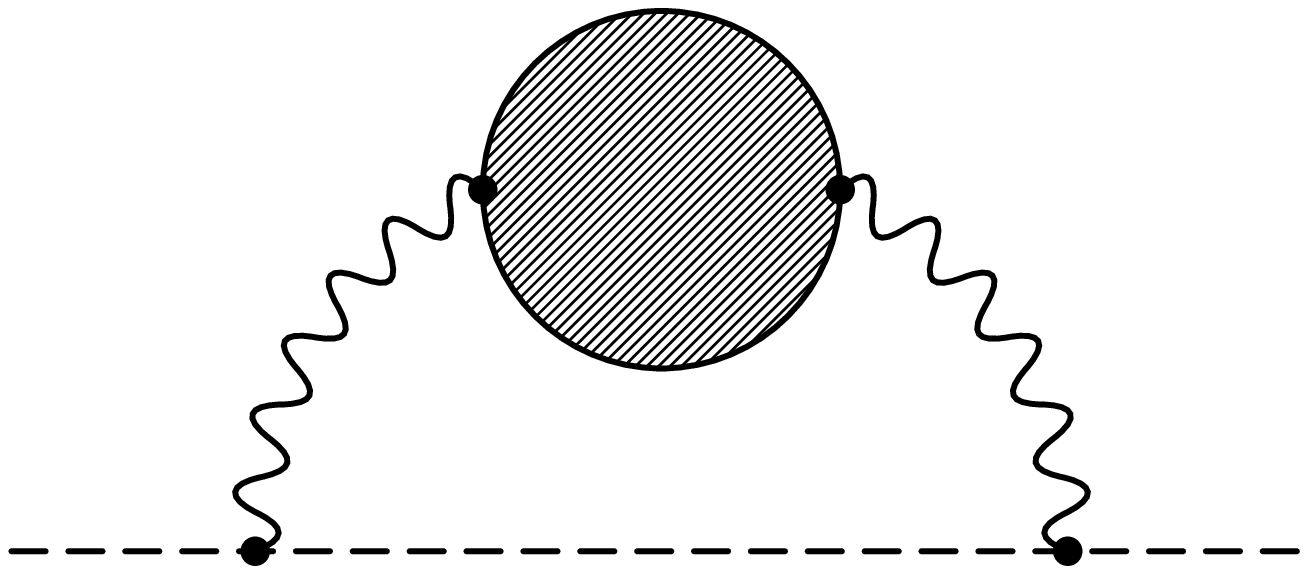}&\epsfxsize=\size truein\epsfbox{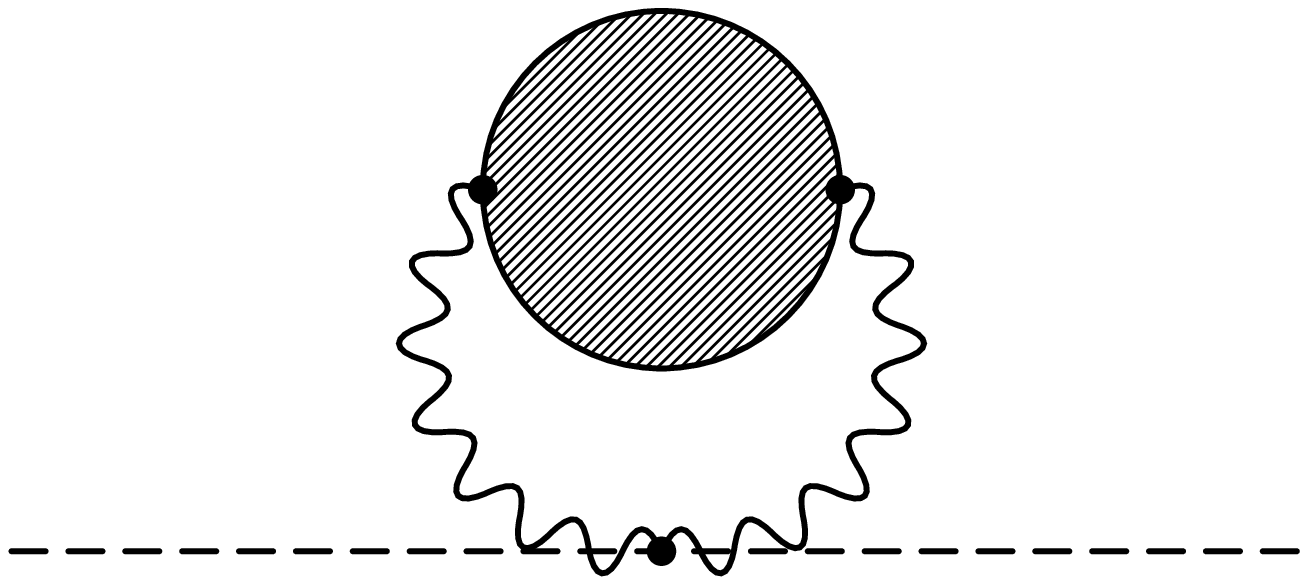}\cr
&\cr
$D4$&$D5$\cr
}
{\small\noindent{\bf Figure 1.}  Diagram $D1$ gives mass to gauginos and is expressible in terms of the function $\wt B_{1/2}$.  Diagrams $D2$-$D5$ contribute to the masses of sfermions and involve the functions $\wt C_0$, $\wt C_{1/2}$, and $\wt C_1$, respectively.}
\vskip.2in
\noindent The divergent part of \ctildejlam\ is the hidden-sector contribution to the gauge beta function:
\eqn\betacontrib{\delta {dg\over d\ln \mu}={g^3\over 16\pi ^2}(2\pi)^4 c,} 
with $c>0$.  

The diagrams of Figure 1, which give masses to the visible sector gauginos and sfermions were evaluated in \mss\ in terms of the current correlator functions as 
\eqn\mssresult{\eqalign{M_a& = g_a^2M\widetilde B_{1/2}^{(a)}(0),\qquad m_{\widetilde f}^2=g_1^2 Y_f \xi + \sum _a g_a^4 c_2(a_f)A_a\cr 
A_a&\equiv -\int {d^4 p \over (2\pi )^4}{1\over p^2}\left(3\widetilde C_1^{(a)}(p^2/M^2)-4\widetilde C_{1/2}^{(a)}(p^2/M^2)+\widetilde C_0^{(a)}(p^2/M^2)\right).}}
The index $a$ runs over the gauge groups, $f$ runs over the sfermions, $Y$ is the hypercharge, and $\xi$ is an FI parameter.   Note that the integrand of $A_a$ has the form of a super-trace\foot{A related expression appears in \dMM\ for the messenger $m_{mess}^2$, in the context of models with a separate messenger sector (where it was argued that perturbative estimates based on naive dimensional analysis should be essentially reliable even for strongly coupled susy-breaking sectors).}  and, without additional information or constraints, it looks like it can have either sign.

\newsec{The effective potential and Higgsed gauge mediation}
\vskip.2in
\centertable{
\vrule height1.8ex depth0.25ex width 0pt \tabskip=1em  \hfil#\hfil\qquad&\qquad\hfil#\hfil\qquad&\qquad\hfil#\hfil\cr
\epsfxsize=1truein\epsfbox{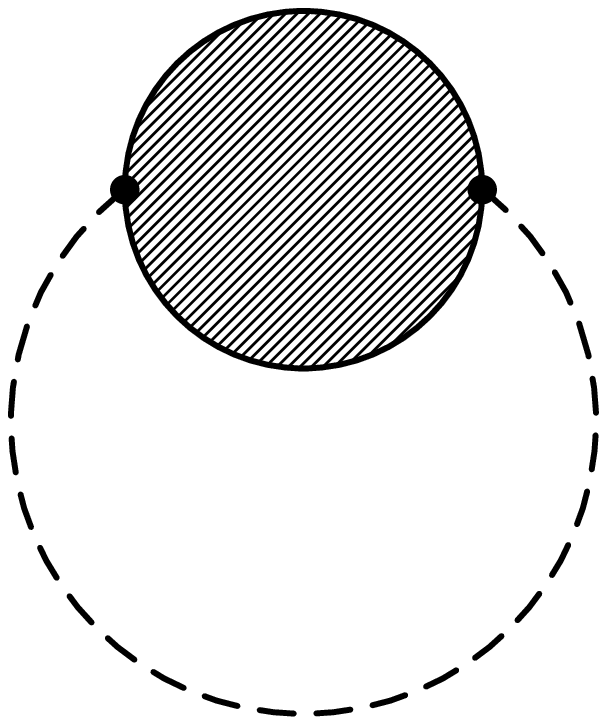}&\epsfxsize=1truein\epsfbox{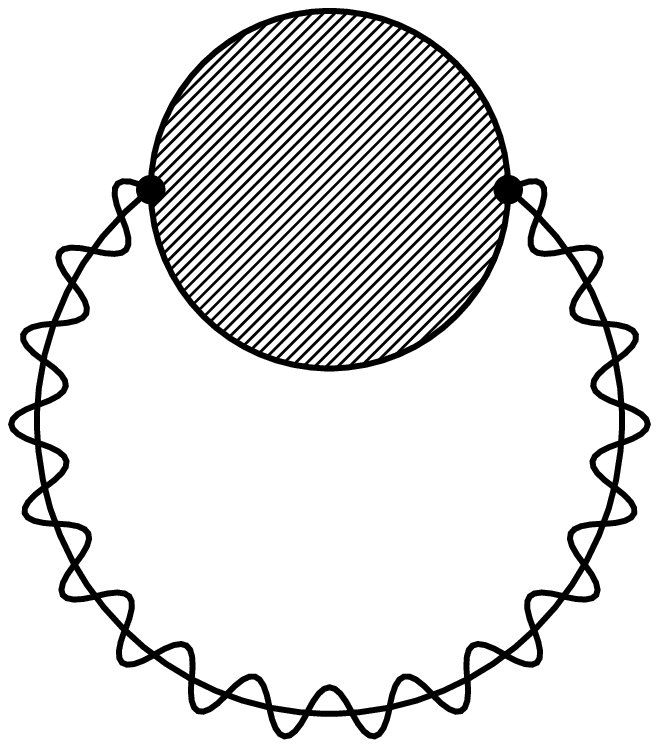}&\epsfxsize=1truein\epsfbox{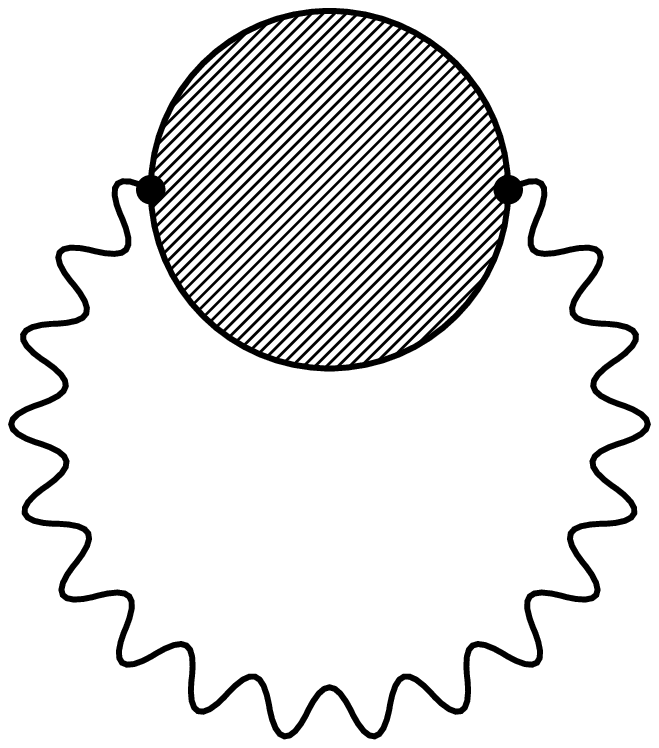}\cr
&\cr
$D6$&$D7$&$D8$\cr
}
{\small\noindent{\bf Figure 2.}  Diagrams $D6$, $D7$, and $D8$ give contributions to the effective potential involving the functions $\wt C_0$, $\wt C_{1/2}$, and $\wt C_1$, respectively.}
\vskip.2in
The sfermions of the visible sector generally have tree-level D-flat directions.  With supersymmetry breaking, these directions are lifted first by the two-loop effective potential.   Near the origin, this effective potential reduces to the sfermion mass terms.  Far from the origin, the effects of susy breaking shut off, and the effective potential becomes very flat.  The full effective potential can be of interest for cosmological models and was computed in \dMM, to leading order in small F-terms, for the case of a weakly coupled messenger sector.  Here we give a simple expression for the full effective potential, for arbitrary F-terms, in general gauge mediation.

The effective potential is computed from the diagrams of Figure 2.   For simplicity, we quote the result for a single $U(1)$ gauge group -- the more general case is similar.    We find
that the effective potential is simply 
\eqn\veffgen{V_{eff}(m_W^2)={g^2 \over 2}\int {d^4 p\over (2\pi )^4}{p^2\over p^2+m_W^2}
\left(3\widetilde C_1(p^2/M^2)-4\widetilde C_{1/2}(p^2/M^2)+ \widetilde C_0(p^2/M^2)\right).}
Here $m_W=2g|\ev{Q}|$ is the mass of the vector multiplet, where $\ev{Q}$ is along a direction which would have been D-flat if not for the supersymmetry breaking.  The terms in \veffgen\ simply come from contracting the massive vector multiplet propagators with the appropriate current-current correlator, e.g. from diagram $D8$ we have (in Euclidean space)
\eqn\cOne{\frac{g^{\mu\nu}}{p^2+m_W^2}(p^2\eta_{\mu\nu}-p_\mu p_\nu)\wt C_1(p^2/M^2)=\frac{p^2}{p^2+m_W^2}3\wt C_1(p^2/M^2).}
Diagram $D7$ is similar, with the massive gaugino propagator contracted with \compjj{b}. 
Diagram $D6$, with the auxiliary field $D$, requires a bit more attention because it mixes with a real scalar, $C$, of the massive gauge multiplet:
\eqn\propCD{\Delta _{CD}=\pmatrix{1&m_W\cr m_W &-p^2}^{-1}={1\over p^2+m_W^2}\pmatrix{p^2 & m_W\cr m_W & -1},}
so the $D$-field propagator is $p^2/(p^2+m_W^2)$, which then yields the $\wt C_0$ contribution to \veffgen.  

Let us now verify that our general effective potential \veffgen\ reduces to the sfermion $m_{\widetilde f}^2$ in \mssresult, when expanded around the origin.  Consider 
\eqn\sfermh{m_{\widetilde f}^2={\partial V_{eff}\over \partial |\ev{Q}|^2}=-2g^4\int {d^4 p\over (2\pi )^4}{p^2\over (p^2+m_W^2)^2}\left(3\widetilde C_1(p^2/M^2)-4\widetilde C_{1/2}(p^2/M^2)+ \widetilde C_0(p^2/M^2)\right).}
Evaluating this for $m_W^2=0$ indeed reduces to the expression of \mss.

One can also verify that the general result \veffgen\ reduces to the effective potential obtained in \dMM, for the special case of weakly coupled hidden sector (using the expressions for $\widetilde C_j$ in the appendix of \mss), when evaluated to leading order in small F-terms.
The case of weakly coupled $\CL _2$ sector, for general (not necessarily small) F-terms, can also be compared and verified with the result obtained using the formulae of \MartinVX.

With no additional work, we can extend our result to allow for the possibility of Higgsed gauge mediation.  We simply replace $m_W^2\to m_W^2+4g^2|\delta Q|^2$ in \veffgen, and keep the $m_W\neq 0$.
Expanding to $\CO (|\delta Q|^2)$, the result \sfermh, with $m_W\neq 0$, gives the sfermion $m_{\widetilde f}^2$ in general Higgsed gauge mediation.  For the case of weakly coupled messengers, it can be verified that the result \sfermh\ indeed reduces to the results obtained in \egms\ for Higgsed gauge mediation.  

\newsec{Superspace techniques and analytic continuation in superspace}
In the context of weakly coupled gauge mediation, there are nice methods \refs{\gr, \aglr}\ which allow multi-loop quantities, including sfermion masses, to be reduced to one-loop quantities at leading order in small supersymmetry breaking F-terms.  Fields $\varphi$ and $\widetilde \varphi$ of the susy-breaking sector get susy-split masses via $W=X\varphi \widetilde \varphi$, where $X$ is a spurion (background) chiral superfield $X=M+\theta ^2 F$. The results follow from imposing the constraints of holomorphy in $X$ on the effective action.  The results are limited to leading order in small $F$, because terms higher order in $F$ arise from higher super-derivative terms in superspace, which are not considered. Taking $x\equiv |F/M^2|\ll 1$, the methods determine the soft masses to $\CO (x)$.  

The methods of \refs{\gr, \aglr}\ extend immediately to general gauge mediation.  The gaugino masses come from the holomorphic gauge coupling $\tau =\theta /2\pi +4\pi i/g^2$, which is a holomorphic function $\tau (X)$ below the scale $X$ thanks to the threshold matching and the contribution \betacontrib\ of the hidden sector to the beta function there.  The sfermion masses come from the one-loop $\CO (g^2)$ contribution to $Z_Q(X, \bar X)$, which depends of $X$ again via the gauge coupling.  This gives $\beta _{g_a}^{(1)}$, and $m^2_{\widetilde f}\sim \gamma ^{(1)}_{\tilde f}\Delta \beta _{g_a}^{(1)}$, 
in terms of the beta-function coefficient $c$ in \betacontrib\ and \ctildejlam:
\eqn\aglrresults{M_a\approx {g_a^2\over 16\pi ^2}(2\pi)^4 c_a{F\over M}, \qquad m_{\widetilde f}^2
\approx\sum_a2c_2(a_f){g_a^4\over (16\pi ^2)^2}(2\pi )^4c_a\Big| {F\over M}\Big| ^2.}
In this small $F$ limit, the masses $m_{\widetilde f}^2$ are manifestly positive.  
 
 The simple expressions \aglrresults\ motivate a parallel spurion analysis of the current correlation functions, to connect with the results of \mss, quoted above in \mssresult, when expanded in small $F$.    In the small-$F$ limit, we have $\wt C_j\approx\wt C_{susy}$, independent of $j=0, 1/2, 1$.  
To leading order in small $F$, we find that  the susy-breaking quantities appearing in the soft masses \mssresult\ can be expressed, for all $p^2/M^2$, as
 \eqn\bidentity{-M\wt B_{1/2}(p^2/M^2)=F\frac{\partial}{\partial M}\wt C_{susy}(p^2/M^2)+\CO(F|F|^2/M^6),}
and 
\eqn\cidentity{3\wt C_1(p^2/M^2)-4\wt C_{1/2}(p^2/M^2)+\wt C_0(p^2/M^2)=2\frac{|F|^2}{p^2}\frac{\partial^2}{|\partial M|^2}\wt C_{susy}(p^2/M^2)+\CO(|F|^4/M^8),}
It is easily verified that these identities are indeed satisfied for the particular case of weakly coupled messengers,  by expanding for small $F/M$ the  explicit expressions for $\wt C_j$ and $\wt B_{1/2}$ in the appendix of \mss.  
The identities \bidentity\ and \cidentity\ were independently derived, with the same motivation, in a recent paper of Distler and Robbins \DistlerBT.

One approach is to prove  \bidentity\ and \cidentity\ directly in terms of the current correlation functions, first enforcing the supersymmetric and current conservation Ward identities, and 
introducing the  supersymmetry
breaking spurion via $M\to X=M+\theta ^2 F$.
The first step is simplified by writing the current supercorrelators in superspace.
In particular, the current 2-point functions for unbroken supersymmetry
are given by an immediate generalization of the conformal result in \os\ to the nonconformal case:
\eqn\superjj{
\eqalign{\vev{\J(z_1)\J(z_2)}
&=\frac{C(M^4x_{\overline21}^2x_{\overline12}^2)}{x_{\overline21}^2x_{\overline12}^2},\qquad x_{\overline12}^\mu=x_{1}^\mu-x_2^\mu-i\theta_1\sigma^\mu\bar\theta_1-i\theta_2\sigma^\mu\bar\theta_2+2i\theta_2\sigma^\mu\bar\theta_1.
\cr
}}
Instead of introducing the spurions in \superjj, we will now do it in terms of the 1PI effective action, since that
is anyway more directly relevant for extracting the implications for gauge mediation.

Consider first the 1PI effective action to $\CO(F^0)$, neglecting supersymmetry breaking effects.  Following  the discussion and notation of \aglr, there is the term involving the gauge fields
\eqn\gammaipi{\eqalign{\Gamma _{1PI}&\supset \int d^4 p \int d^4 \theta\  \gamma (p^2)W^\alpha {D^2\over -8p^2}W_\alpha +h.c.\cr
&=\int d^4 p \int d^2\theta \half \gamma (p^2) W^\alpha W_\alpha +h.c..}}
There are also terms involving the visible sector matter, e.g.  
\eqn\zetaipi{\Gamma _{1PI}\supset \int d^4 p\int d^4\theta \zeta (p^2)\big(Q^\dagger e^VQ+\wt Q^\dagger e^{-V}\wt Q\big).}   
Classically, $\gamma (p^2)=1/g^2$, and one can define the quantum running couplings and $Z$-factors as $\gamma (p^2)|_{p^2=-\mu ^2}=1/g^2(\mu ^2)$, and $\zeta_f (p^2)|_{p^2=-\mu^2}=Z_f(\mu^2)$.
To $\CO (g^2)$, the hidden sector current-current two-point functions contribute 
\eqn\candgamma{\delta \gamma(p^2/M^2)=g^2\wt C_{susy}(p^2/M^2).}
The leading term in the low-momentum expansion of \gammaipi\ then gives the terms \gaugecont\ in the effective Lagrangian (in the susy limit).   We are here interested in the relation \candgamma\ for general $p^2$.  

We now introduce the supersymmetry breaking spurion via $M\to X=M+\theta ^2F$.  To $\CO(|F|^2)$, the relation \candgamma\ is simply preserved, and both sides pick up $\theta$ components.  The 1PI action then includes
\eqn\gammaipii{\eqalign{\Gamma _{1PI}&\supset \half\int d^4 p \int d^2 \theta\  \big(\gamma (p^2)|_0+\theta ^2\gamma (p^2)|_{\theta^2} \big)W^\alpha W_\alpha  +h.c.\cr
&+\int d^4p\ \gamma (p^2)|_{\theta ^2 \bar \theta ^2}{\lambda \sigma ^\mu p_\mu \overline\lambda \over -p^2}}}
where, according to \candgamma\ with the spurions, we have 
\eqn\gammacomps{\gamma (p^2)|_{\theta ^2}= g^2F{\partial\over \partial X} \widetilde C_{susy}(p^2/|X|^2)\bigg|_{X=M},\qquad \gamma (p^2)_{\theta ^4}=g^2 |F|^2{\partial ^2\over |\partial X|^2}\widetilde C_{susy}(p^2/|X|^2)\bigg|_{X=M}.}
The  $\gamma (p^2)|_{\theta ^2}$ term in \candgamma\ corresponds  to the non-supersymmetric analog of \candgamma, generated in the effective action by the non-supersymmetric current two-point functions:
\eqn\gammathetais{\gamma (p^2)|_{\theta ^2}=-g^2M\widetilde B_{1/2}(p^2/M^2) - \CO(F|F|^2),}
where on the RHS we keep only the $\CO(F)$ term, and the factor of $-M$ is as in \gaugecont.  Comparing \gammacomps\ and \gammathetais\ gives the relation \bidentity. 

Likewise, the $\gamma (p^2)|_{\theta ^4}$ term in \gammaipii\ is generated by the non-supersymmetric analog of \candgamma, from supersymmetry breaking contributions of the current-current two-point functions to the effective action.  Indeed, it is easily seen, as in \gaugecont,  that a supersymmetry breaking shift of $\wt C_{1/2}(p^2/M^2)$ will generate the $\gamma (p^2)|_{\theta ^4}$ term in \gammaipii.   Since this supersymmetry breaking term comes from such a $\wt C_{1/2}$ shift relative to $\wt C_0$ and $\wt C_{1}$, it is generated only by the supertrace:
\eqn\gammathetaiv{{1\over p^2}\gamma (p^2)|_{\theta ^4}=\frac{g^2}{2} \left(\wt C_0(p^2/M^2)-4\wt C_{1/2}(p^2/M^2)+3\wt C_1(p^2/M^2)\right)-\CO(|F|^4)),}
where only terms to $\CO (|F|^2)$ are kept on the RHS.  
This relation essentially appears already in \aglr\ (in terms of the gauge field effective propagators), where it was noted to follow from considering all the possible contributing terms  involving the spurion and supercovariant derivatives acting on it.   Comparing \gammathetaiv\ with \gammacomps\ yields the relation \cidentity.  

It is evident that the $\theta ^2$ term in \gammaipii\ yields the gaugino mass, and \gammathetais\ agrees with the result of \mss, quoted above in \mssresult.  This result for $M_a$ indeed agrees with \aglrresults, as seen from \bidentity\ and the contribution of the $\log \Lambda$ term in \ctildejlam\   to $\wt C_{susy}$.    The expression for $m_{\tilde f}^2$ quoted above in \aglrresults\ follows from \cidentity\ and the result \mssresult.  

\bigskip

\noindent {\bf Acknowledgments:}

This research was supported in part by UCSD
grant DOE-FG03-97ER40546.  KI would like to thank the International Centre for Theoretical Studies of the Tata Institute for hospitality and support during the time when this paper was written up.

\listrefs\end